\documentclass[conference]{IEEEtran}
\IEEEoverridecommandlockouts

\usepackage{cite}
\usepackage{amsmath,amssymb,amsfonts}
\usepackage{algorithmic}
\usepackage{graphicx}
\usepackage{textcomp}
\usepackage{xcolor}
\def\BibTeX{{\rm B\kern-.05em{\sc i\kern-.025em b}\kern-.08em
    T\kern-.1667em\lower.7ex\hbox{E}\kern-.125emX}}

\usepackage{siunitx}				
\usepackage{hyperref}
\usepackage{booktabs}  
\newcommand{\ra}[1]{\renewcommand{\arraystretch}{#1}}
\usepackage{tikz}
\usepackage{tikz-3dplot}
\usepackage{pgfplots}
\pgfplotsset{compat=newest}
\newlength\figureheight 
\newlength\figurewidth 
\usepackage{subcaption}
\usepackage{scalerel}
\usepackage{tikz}
\usetikzlibrary{svg.path}

\definecolor{orcidlogocol}{HTML}{A6CE39}
\tikzset{
  orcidlogo/.pic={
    \fill[orcidlogocol] svg{M256,128c0,70.7-57.3,128-128,128C57.3,256,0,198.7,0,128C0,57.3,57.3,0,128,0C198.7,0,256,57.3,256,128z};
    \fill[white] svg{M86.3,186.2H70.9V79.1h15.4v48.4V186.2z}
                 svg{M108.9,79.1h41.6c39.6,0,57,28.3,57,53.6c0,27.5-21.5,53.6-56.8,53.6h-41.8V79.1z M124.3,172.4h24.5c34.9,0,42.9-26.5,42.9-39.7c0-21.5-13.7-39.7-43.7-39.7h-23.7V172.4z}
                 svg{M88.7,56.8c0,5.5-4.5,10.1-10.1,10.1c-5.6,0-10.1-4.6-10.1-10.1c0-5.6,4.5-10.1,10.1-10.1C84.2,46.7,88.7,51.3,88.7,56.8z};
  }
}

\newcommand\orcidicon[1]{\href{https://orcid.org/#1}{\mbox{\scalerel*{
\begin{tikzpicture}[yscale=-1,transform shape]
\pic{orcidlogo};
\end{tikzpicture}
}{|}}}}

\newcommand*\diff{\mathop{}\!\mathrm{d}}

\makeatletter
\newcommand{\fcdot}{\,\cdot\,}
\newcommand{\fcarg}[1]{\def\fc@rg{#1}\ifx\fc@rg\empty\fcdot\else\fc@rg\fi}
\makeatother

\newcommand{\absfrac}{\absfrac}

\newcommand{\norm}[2][]{\lVert\fcarg{#2}\rVert\ifx#1\empty\else_{#1}\fi}
\newcommand{\Norm}[2][]{\left\lVert#2\right\rVert\ifx#1\empty\else_{#1}\fi}

\newcommand{\snorm}[2][]{\lvert\!\lvert\!\lvert
  \fcarg{#1}\rvert\!\rvert\!\rvert\ifx#2\empty\else_{#1}\fi}
\newcommand{\Snorm}[2][]{\left\lvert\!\left\lvert\!\left\lvert
  #2\right\rvert\!\right\rvert\!\right\rvert\ifx#1\empty\else_{#1}\fi}
\newcommand{\sprod}[3][]{%
  \langle\fcarg{#2},\fcarg{#3}\rangle\ifx#1\empty\else_{#1}\fi}
\newcommand{\Sprod}[3][]{%
  \left\langle\fcarg{#2},\fcarg{#3}\right\rangle\ifx#1\empty\else_{#1}\fi}

\newcommand{\optmathindex}[1]{\ifx#1\empty\else,#1\fi}
\newcommand{\opttextindex}[1]{\ifx#1\empty\else,\text{#1}\fi}
\newcommand{\optmathsb}[1]{\ifx#1\empty\else_{#1}\fi}
\newcommand{\opttextsb}[1]{\ifx#1\empty\else_{\text{#1}}\fi}
\newcommand{\optmathsp}[1]{\ifx#1\empty\else^{#1}\fi}
\newcommand{\opttextsp}[1]{\ifx#1\empty\else^{\text{#1}}\fi}

\newcommand{\continuousFunctions}[1]{\mathcal{C}\ifx#1\empty\else^{#1}\fi}

\newcommand{\piecewiseContinuousFunctions}[1]{\mathcal{C}_p\ifx#1\empty\else^{#1}\fi}

\newcommand{\define}{\mathrel{{\mathop:}{=}}}

\hypersetup{pdfauthor={Nils Poschadel, Robert Hupke, Stephan Preihs and J{\"u}rgen Peissig},
            pdftitle={Direction of Arrival Estimation of Noisy Speech using Convolutional Recurrent Neural Networks with Higher-Order Ambisonics Signals},
            pdfsubject={EUSIPCO 2021},
            pdfkeywords={DOA, HOA, FOA, CRNN},
            pdfborder = {0 0 0}
            }

\begin{document}

\title{Direction of Arrival Estimation of Noisy Speech using Convolutional Recurrent Neural Networks with Higher-Order Ambisonics Signals}
\author{
\IEEEauthorblockN{Nils Poschadel \orcidicon{0000-0003-4302-5569}\,, Robert Hupke \orcidicon{0000-0002-8304-8610}\,, Stephan Preihs \orcidicon{0000-0003-4232-580X}\,, and J{\"u}rgen Peissig \orcidicon{0000-0002-4649-8911}}
\IEEEauthorblockA{Institute of Communications Technology\\
Leibniz University Hannover, Hannover, Germany\\Email: \{poschadel, hupke, preihs, peissig\}@ikt.uni-hannover.de}
}

\maketitle

\begin{abstract}
Training convolutional recurrent neural networks on first-order Ambisonics signals is a well-known approach when estimating the direction of arrival for speech/sound signals. In this work, we investigate whether increasing the order of Ambisonics up to the fourth order further improves the estimation performance of convolutional recurrent neural networks. While our results on data based on simulated spatial room impulse responses show that the use of higher Ambisonics orders does have the potential to provide better localization results, no further improvement was shown on data based on real spatial room impulse responses from order two onwards. Rather, it seems to be crucial to extract meaningful features from the raw data. First order features derived from the acoustic intensity vector were superior to pure higher-order magnitude and phase features in almost all scenarios.

\end{abstract}

\begin{IEEEkeywords}
Direction of arrival estimation, higher-order Ambisonics, convolutional recurrent neural network, spherical harmonics.
\end{IEEEkeywords}

\section{Introduction}
\label{sec:intro}
 Estimating the direction of arrival (DOA) of sound/speech is a key problem in acoustic signal processing. Neural networks have been shown to be superior to classical parametric approaches in this task, especially in reverberant, noisy, and low-SNR environments \cite{Chakrabarty.2019, Xiao.2015, Adavanne.2019, Perotin.2019}.
Recently, DOA estimation based on first-order Ambisonics (FOA) signals has been the subject of much attention \cite{Perotin.2019, Kapka.2019, Politis.06.09.2020, Cao.2019}. Due to the flexibility and generalizability of the Ambisonics approach, it  more or less enables microphone-array-independent DOA estimation models.

Perotin et al. \cite{Perotin.2019, Perotin.2018, Perotin.2019b} investigated the effect of different parameters when training convolutional recurrent neural networks (CRNNs) on FOA data for the DOA estimation of noisy speech. They proposed the usage of features derived from the sound intensity vector as input for the training, achieving greater accuracy in DOA estimation than with using pure magnitude and phase information \cite{Perotin.2018}. 
Furthermore, they showed that a regression approach is at least as suitable as a classification interpretation for a single-source DOA estimation with diffuse interference and that a CRNN trained on spherical coordinates performs worse than a network trained on Cartesian coordinates when using the mean squared error (MSE) or angular distance as loss function \cite{Perotin.2019b}.

However, despite the increasing availability of higher-order ambisonics (HOA) microphones, 
very little research is conducted on the performance of DOA estimators based on HOA signals. 

There are some results from other applications where the usage of HOA signals is advantageous over the use of FOA signals. Pointer experiments with subjects showed a positive influence of the order of the Ambisonics signal on perceptual localization accuracy in a loudspeaker reproduction of a sound field \cite{Bertet.2013}.
Similarly, the higher-order model of directional audio coding (HO-DirAC) achieved a higher reproduction accuracy than the first-order DirAC in a perceptual evaluation \cite{Politis.2015, Pulkki.2018b}.
Investigations on spherical harmonic (SH) beamforming with unsupervised peak clustering \cite{Green.25.10.201926.10.2019} also showed an improvement in localization accuracy with increasing Ambisonics order. However, to our knowledge, this topic has not yet been investigated or even quantified for state of the art deep learning approaches for DOA estimation. 

This work therefore is the first to apply the idea of CRNN-based DOA estimation to HOA signals and to investigate whether or how much the additional spatial information contained in HOA can improve the estimation accuracy.
We thereby compare our HOA models with FOA models based on both magnitude/phase spectrograms and spectrogram features derived from the acoustic intensity vector.

To the best of our knowledge, there is no sufficiently large dataset of HOA speech signals or impulse responses available. Therefore, we had to create a suitable dataset of noisy speech data with different orders of Ambisonics, taking inspiration from the procedure used in \cite{Perotin.2019b} for creating a FOA dataset. Due to the way we parameterize the impulse response simulation, this dataset can not only be used for training deep learning models for DOA estimation. The dataset also contains labels regarding room size/geometry as well as acoustic properties such as reverberation time and absorption/scattering coefficients and will serve as the basis for a number of studies in the context of acoustic analysis based on HOA signals.

We present the details on the generation of our training, validation, and testing data in Sec. \ref{sec:data} after a brief introduction to the fundamentals of Ambisonics and SH in Sec. \ref{sec:ambisonics}. The configuration of the trained model and the metrics are described in Sec. \ref{sec:doa_framework}. Finally, the results based on simulated and measured data are compared and discussed in Sec. \ref{sec:results} and summarized in Sec. \ref{sec:conclusion}.

\section{Ambisonics}
\label{sec:ambisonics}
Ambisonics is a 3D audio surround representation and rendering approach based on the spatial decomposition of the sound field in the orthonormal basis of SH \cite{Nachbar.2011, Perotin.2019}. This section gives an overview of the mathematical principles of Ambisonics. This condensed description of the
SH decomposition is based on the more detailed presentation in \cite{Rafaely.2015, Zotter.2019}. 

In the following, the Cartesian $(x, y, z)\in\mathbb{R}^3$ and the spherical $(r, \theta, \phi) = (r, \Omega)\in [0,\infty)\times (-\frac{\pi}{2},\frac{\pi}{2}]\times[-\pi, \pi]$ coordinate systems are used.
The $x$-, $y$-  and $z$-axes point to the front, left and top, respectively. The angle $\phi$ is the azimuth, which is zero at the frontal direction and increasing counterclockwise; $\theta$ is the elevation, which is zero at the
horizontal plane and positive above, and $r$ is the radius.

Consider a function $f(\theta, \phi) = f(\Omega)\in\ L^2\left(S^2\right)$ on the unit 2-sphere $S^2 \define \left\{\mathbf{x}\in\mathbb{R}^3:\norm{\mathbf{x}}_2=1\right\}$, then the SH decomposition of $f$ is given by

\begin{equation}\label{eq:inv_sft}
    f(\Omega) = \sum_{n=0}^\infty \sum_{m=-n}^n f_{nm}{Y_n^m}(\Omega),
\end{equation}
where $Y_n^m$ is the \emph{spherical harmonic} of order~$n$
and degree~$m$.
The coefficients $f_{nm}$ are calculated by

\begin{equation}\label{eq:sft}
    f_{nm} = \int_{\Omega \in S^2} f(\Omega) {Y_n^m}^*(\Omega) \diff \Omega, 
\end{equation}
where $\int_{\Omega \in S^2} \diff \Omega = \int_{-\pi}^{\pi}\int_{-\pi/2}^{\pi/2}  \sin \theta \diff \theta\diff\phi$.
Equations (\ref{eq:inv_sft}) and (\ref{eq:sft}) show that any square-integrable function
on the unit 2-sphere can be approximated by a linear combination of the SH. This approximation even becomes exact for an infinite number of SH.
In this paper, the ambiX format \cite{Nachbar.2011} is used for the (real) SH $Y_n^m$:
\begin{equation*}
     {Y_n^m}(\theta, \phi) = {N_n^{\vert m\vert}}  {P_n^{\vert m\vert}}(\sin(\theta))
     \begin{cases}
     \sin(\vert m\vert \phi), \text{for } m < 0\\
     \cos(\vert m\vert \phi), \text{for } m \geq 0
     \end{cases}
\end{equation*}
with the Legendre-functions $P_n^m$.
To build the set of Ambisonics signals according to ambiX, the channels corresponding
to the SH are ordered by the Ambisonics channel number $\text{ACN}= n^2 + n + m$
and normalised by the SN3D normalisation 
\begin{equation*}
    N_n^{\vert m\vert} = \sqrt{\frac{2-\delta_m}{4\pi}\frac{(n-\vert m\vert)!}{(n+\vert m\vert)!}}.
\end{equation*}
In the special case of FOA, the channels 1-4 according to ACN are often referred to as $W,Y,Z,X$.


\section{Data}
\label{sec:data}
\subsection{Simulated SRIRs}
The training, validation and testing data was generated from a set of spatial room impulse responses (SRIRs) simulated with the MCRoomSim toolbox \cite{Wabnitz.2010} as Ambisonics signals up to fourth order corresponding to the ambiX format. The approach was inspired by the procedure described in \cite{Perotin.2019}. Alltogether we generated \num{8000}, \num{500}, and \num{500} rooms with random dimensions in $\left[3, 20\right]\times\left[3, 20\right]\times\left[3,5\right]$\,\si{m} for the training, validation, and testing set, respectively. The acoustic properties of the walls (frequency dependant scattering and absorption coefficients) were set to plausible, randomly chosen surfaces of the GRAP database \cite{Ackermann.2018}.
For every room, one receiver was randomly positioned with a minimum distance of \SI{1.5}{m} to the walls. Furthermore, one source was randomly positioned at 8 different locations such that the DOAs in the dataset are uniformly distributed. The distance from the source to the receiver was chosen randomly, ensuring that the source and the receiver are at least \SI{1}{m} apart from each other and that the source is at least \SI{49}{\cm} from a wall. With this setup, we simulated \num{64000}, \num{4000} and \num{4000} fourth-order Ambisonics SRIRs. Although the experiments in this paper were conducted using speech signals with a sampling rate of \SI{16}{\kHz}, the SRIRs were simulated with a sampling frequency of \SI{48}{\kHz} to be able to expand the methods of this paper to general audio/music signals using the same database.
After resampling, the SRIRs were convolved with a randomly chosen sentence from the TIMIT database \cite{Garofolo.1993}. This database contains a total of \num{6300} sentences, 10 sentences spoken by each of the 630 speakers (192 female, 438 male) from eight major dialect regions of the United States.
The TIMIT database was split into training, testing and validation sets resulting in 462 (136 female, 326 male), 88 (30 female, 58 male), and 80 (26 female, 54 male) speakers, respectively. 
The training set corresponds to the recommended one by the authors of the TIMIT database. The test set includes the recommended core test set and it is ensured that there is at least one female/male speaker per dialect in the validation and test set, respectively.


Furthermore, we added ambient noise to the speech signals similar to the procedure in \cite{Perotin.2019}. Therefore, we generated single-channel babble noise by overlaying 50 sentences of the respective sets. This babble noise was then convolved with a diffuse SRIR, which was generated by averaging three simulated diffuse parts of SRIRs with a receiver placed in the middle of a random room and a randomly positioned source. This ambient noise was added to the speech signal at a  signal-to-noise ratio (SNR) between 0 and \SI{20}{dB}. Finally, these sentences were cut to one-second-sequences which led to \num{164303}, \num{10285} and \num{10394} sequences for the training, validation and testing set, respectively.

\subsection{Real SRIRs}
For the analysis of DOA estimation performance in a more realistic scenario, we measured real SRIRs in the Immersive Media Lab (IML) \cite{Hupke.2018} at the Institute of Communications Technology. We measured the SRIRs from each of our 36 KH120 loudspeakers to an em32 Eigenmike\textsuperscript{\textregistered} \cite{Meyer.2002} 
microphone at nine different positions, each with two different heights and eight different orientations of our microphone.
In total, the described procedure led to \num{5184} measured SRIRs in the IML, which were afterwards encoded to a fourth-order Ambisonics signal using the EigenUnit-em32-encoder\footnote{https://mhacoustics.com/eigenunits}. These measured SRIRs were used according to the same procedure as for the simulated SRIRs to generate HOA multispeaker signals which resulted in \num{13414} sequences for the testing set based on real SRIRs.

\section{Doa Estimation Framework}
\label{sec:doa_framework}

\subsection{Networks and metrics}
\label{sec:networks}
Our trained networks follow a similar basic CRNN structure compared to the ones in \cite{Adavanne.2019, Perotin.2019}.
A detailed overview of the network's architecture is given in Table~\ref{tab:crnn}, where the final normalization layer scales the prediction to lie on the unit 2-sphere.
We formulated this task as a regression problem with the MSE loss function and the Nadam optimizer \cite{Dozat.2015}.
For training the network, we used the TensorFlow platform \cite{Abadi.2015}.
\begin{table}[htb]\centering
 	\ra{0.85}
    \begin{tabular}{@{}lcrcr@{}}\toprule[1pt]
    	\textbf{Layer} & \phantom{} &\textbf{Details} & \phantom{} &\textbf{Output Shape}\\ \midrule
    	Input && Spectrograms && (50, 512, $dim_\text{in}$) \\ 
    	Conv2D \rule{0pt}{2.5ex}&& $3\times 3$ && (50, 512, $n_\text{filter}$) \\	
    	BatchNorm &&  && (50, 512, $n_\text{filter}$) \\
    	Activation && elu && (50, 512, $n_\text{filter}$) \\
    	MaxPooling && $1\times 8$ && (50, 64, $n_\text{filter}$) \\
    	Dropout && 0.2 && (50, 64, $n_\text{filter}$) \\
    	Conv2D \rule{0pt}{2.5ex}&& $3\times 3$ && (50, 64, $n_\text{filter}$) \\	
    	BatchNorm &&  && (50, 64, $n_\text{filter}$) \\
    	Activation && elu && (50, 64, $n_\text{filter}$) \\
    	MaxPooling && $1\times 8$ && (50, 8, $n_\text{filter}$) \\
    	Dropout && 0.2 && (50, 8, $n_\text{filter}$) \\
    	Conv2D \rule{0pt}{2.5ex}&& $3\times 3$ && (50, 8, $n_\text{filter}$) \\	
    	BatchNorm &&  && (50, 8, $n_\text{filter}$) \\
    	Activation && elu && (50, 8, $n_\text{filter}$) \\
    	MaxPooling && $1\times 4$ && (50, 2, $n_\text{filter}$) \\
    	Dropout && 0.2 && (50, 2, $n_\text{filter}$) \\
    	Reshape \rule{0pt}{2.5ex}&&   && (50, 2$\cdot n_\text{filter}$) \\
    	BiLSTM &&   && (50, 2$\cdot n_\text{filter}$) \\
    	BiLSTM &&   && (50, 2$\cdot n_\text{filter}$) \\
    	Time-Dist. Dense \rule{0pt}{2.5ex}&& elu  && (50, 2$\cdot n_\text{filter}$) \\
    	Dropout && 0.2 && (50, 2$\cdot n_\text{filter}$) \\
    	Time-Dist. Dense && linear && (50, 3) \\
    	Normalization && && (50, 3) \\
    	\bottomrule[1pt] 
    \end{tabular} 
    \caption{Architecture of the CRNNs for DOA estimation.}
    \label{tab:crnn}
\end{table}
Since we use a time-distributed output layer and assume the sources to be static over the whole duration of the signal, we first average the network outputs for each axis over time. We then compare the predicted DOA $(\hat{\theta},\hat{\phi})$ with the reference $(\theta,\phi)$ used to synthesize the dataset, using the \mbox{\emph{angular distance}} $\delta\bigl[(\hat{\theta},\hat{\phi}),(\theta,\phi)\bigr]$ defined by
\begin{align*}\label{eq:angular_dist}
    \delta\bigl[(\hat{\theta},\hat{\phi}),(\theta,\phi)\bigr]  = \arccos&\bigl[\sin{(\hat{\theta})}\sin{(\theta)}
    \\&+ \cos{(\hat{\theta})}\cos{(\theta)}\cos{(\hat{\phi}-\phi)}\bigr].
\end{align*}
For additional evaluation, we further define the so-called \emph{accuracy} as the proportion of samples for which the prediction has an angular distance below a given error tolerance.

\subsection{Input features}
\label{sec:input_features}
The input features of the networks based on HOA signals are pure magnitude and phase spectrograms. In the following, we will call these networks HOA-$n$-CRNN with $n$ being the order of the HOA signal. We compare our HOA-$n$-CRNNs to two other published approaches for FOA DOA estimation with CRNNs. On the one hand, Adavanne et al. \cite{Adavanne.2019} used pure FOA magnitude and phase spectrograms (FOA-CRNN). Of course, HOA-1-CRNN and FOA-CRNN are identical and will be referred to as HOA-1-CRNN in the following. On the other hand, Perotin et al. \cite{Perotin.2019} proposed using spectrograms of 6-channel features derived from the FOA sound intensity vector according to (\ref{eq:intensity}) as input to the CRNN (Intensity-CRNN). By using these features, they were able to significantly improve the localization performance compared to using magnitude and phase spectrograms.
\begin{equation}\label{eq:intensity}
    \frac{-1}{C(t,f)}\begin{bmatrix} \mathbf{I}_a(t,f) \\ \mathbf{I}_r(t,f)\end{bmatrix}
\end{equation}
$\mathbf{I}_a(t,f)$ and $\mathbf{I}_r(t,f)$ describe the active and reactive intensity vector as a Short-time Fourier transform (STFT) expression of the FOA channels and $C(t,f)$ is a normalization term. They can be computed according to
(\ref{eq:act_int}), (\ref{eq:react_int}), (\ref{eq:norm_int}).
For further details on acoustic intensity see \cite{Perotin.2019,Pulkki.2018,Jacobsen.1991}.

\begin{equation}\label{eq:act_int}
    \mathbf{I}_a(t,f) = -
        \begin{bmatrix}
            \operatorname{Re}\left\{W(t,f)X^*(t,f)\right\} \\
            \operatorname{Re}\left\{W(t,f)Y^*(t,f)\right\} \\
            \operatorname{Re}\left\{W(t,f)Z^*(t,f)\right\} \\
        \end{bmatrix}
\end{equation}
\begin{equation}\label{eq:react_int}
    \mathbf{I}_r(t,f) = -
        \begin{bmatrix}
            \operatorname{Im}\left\{W(t,f)X^*(t,f)\right\} \\
            \operatorname{Im}\left\{W(t,f)Y^*(t,f)\right\} \\
            \operatorname{Im}\left\{W(t,f)Z^*(t,f)\right\} \\
        \end{bmatrix}
\end{equation}
\begin{equation}
\label{eq:norm_int}
    C(t,f) = \lvert W(t,f)\rvert^2+ \frac{1}{3}(\lvert X(t,f)\rvert^2+\lvert Y(t,f)\rvert^2+\lvert Z(t,f)\rvert^2)
\end{equation}
The input shape of all the different networks is $(50, 512, dim_\text{in})$, 
where  \num{50} is the number of frames,  \num{512} the number of frequency bins, and $dim_\text{in}$ the number of input channels with $dim_\text{in} = 2(n+1)^2$ for the HOA-$n$-CRNNs and $dim_\text{in} = 6$ for the Intensity-CRNN.
The STFT for the creation of the spectrograms was performed on \num{640} samples, zero-padded to \num{1024} samples with a hop-size of  \num{320} samples.
For identifying the optimal number of filters ($n_\text{filter}$), different values ranging from  \num{32} to \num{1024} were tested for each network
and the value which resulted in the lowest error on the validation set was chosen. The best values were \num{256} for the HOA-1-CRNN and HOA-2-CRNN and \num{512} for all the other networks.


\section{Results}
\label{sec:results}
\setlength\figureheight{5.5cm} 
\begin{figure*}[htb]
    \begin{subfigure}[htb]{0.49\linewidth}
        \caption{Simulated SRIRs.}\label{fig:box_sim}
            \vspace*{-10pt}
    \begin{center}
        \begin{small}
\begin{tikzpicture}

\begin{axis}[
width=\figurewidth,
height=\figureheight,
at={(0\figurewidth,0\figureheight)},
tick align=outside,
tick pos=left,
x grid style={white!80!black},
xlabel={Network type},
xmin=0.5, xmax=5.5,
xtick style={color=black},
xtick={1,2,3,4,5},
xticklabels={Intensity,HOA-1,HOA-2,HOA-3,HOA-4},
y grid style={white!80!black},
ymajorgrids,
ylabel={Angular distance / $^\circ$},
ymin=-0.5, ymax=14,
ytick={0,2,4,6,8,10,12,14},
ytick style={color=black}
]
\addplot [black]
table {%
0.75 1.0779822495168
1.25 1.0779822495168
1.25 2.71129031931656
0.75 2.71129031931656
0.75 1.0779822495168
};
\addplot [black]
table {%
1 1.0779822495168
1 0.0158581219084665
};
\addplot [black]
table {%
1 2.71129031931656
1 5.15230329374381
};
\addplot [black]
table {%
0.875 0.0158581219084665
1.125 0.0158581219084665
};
\addplot [black]
table {%
0.875 5.15230329374381
1.125 5.15230329374381
};
\addplot [black]
table {%
1.75 1.65708205893487
2.25 1.65708205893487
2.25 4.22911881985459
1.75 4.22911881985459
1.75 1.65708205893487
};
\addplot [black]
table {%
2 1.65708205893487
2 0.0249778338841183
};
\addplot [black]
table {%
2 4.22911881985459
2 8.08486760508328
};
\addplot [black]
table {%
1.875 0.0249778338841183
2.125 0.0249778338841183
};
\addplot [black]
table {%
1.875 8.08486760508328
2.125 8.08486760508328
};
\addplot [black]
table {%
2.75 1.56040726191569
3.25 1.56040726191569
3.25 3.80894545145463
2.75 3.80894545145463
2.75 1.56040726191569
};
\addplot [black]
table {%
3 1.56040726191569
3 0.0235604049021846
};
\addplot [black]
table {%
3 3.80894545145463
3 7.17398534957437
};
\addplot [black]
table {%
2.875 0.0235604049021846
3.125 0.0235604049021846
};
\addplot [black]
table {%
2.875 7.17398534957437
3.125 7.17398534957437
};
\addplot [black]
table {%
3.75 1.26802360183164
4.25 1.26802360183164
4.25 3.09327245260925
3.75 3.09327245260925
3.75 1.26802360183164
};
\addplot [black]
table {%
4 1.26802360183164
4 0.0291584093068258
};
\addplot [black]
table {%
4 3.09327245260925
4 5.82981719925825
};
\addplot [black]
table {%
3.875 0.0291584093068258
4.125 0.0291584093068258
};
\addplot [black]
table {%
3.875 5.82981719925825
4.125 5.82981719925825
};
\addplot [black]
table {%
4.75 1.18944757148857
5.25 1.18944757148857
5.25 2.85035517284183
4.75 2.85035517284183
4.75 1.18944757148857
};
\addplot [black]
table {%
5 1.18944757148857
5 0.0264930411091193
};
\addplot [black]
table {%
5 2.85035517284183
5 5.34159364406805
};
\addplot [black]
table {%
4.875 0.0264930411091193
5.125 0.0264930411091193
};
\addplot [black]
table {%
4.875 5.34159364406805
5.125 5.34159364406805
};
\addplot [black]
table {%
0.75 1.76937647310791
1.25 1.76937647310791
};
\addplot [black]
table {%
1.75 2.69561034409807
2.25 2.69561034409807
};
\addplot [black]
table {%
2.75 2.51601550093894
3.25 2.51601550093894
};
\addplot [black]
table {%
3.75 2.04879243757528
4.25 2.04879243757528
};
\addplot [black]
table {%
4.75 1.91381969961821
5.25 1.91381969961821
};
\end{axis}

\end{tikzpicture}
        \end{small}
    \end{center}
    \end{subfigure}
    \begin{subfigure}[htb]{0.49\linewidth}
    \caption{Real SRIRs.}\label{fig:box_real}
    \vspace*{-10pt}
    \begin{center}
        \begin{small}
\begin{tikzpicture}

\begin{axis}[
width=\figurewidth,
height=\figureheight,
at={(0\figurewidth,0\figureheight)},
tick align=outside,
tick pos=left,
x grid style={white!80!black},
xlabel={Network type},
xmin=0.5, xmax=5.5,
xtick style={color=black},
ymajorgrids,
xtick={1,2,3,4,5},
xticklabels={Intensity,HOA-1,HOA-2,HOA-3,HOA-4},
y grid style={white!80!black},
ylabel={Angular distance / $^\circ$},
ymin=-0.5, ymax=14,
ytick={0,2,4,6,8,10,12,14},
ytick style={color=black}
]
\addplot [black]
table {%
0.75 2.69017845511818
1.25 2.69017845511818
1.25 6.04109586225809
0.75 6.04109586225809
0.75 2.69017845511818
};
\addplot [black]
table {%
1 2.69017845511818
1 0.0142897494772524
};
\addplot [black]
table {%
1 6.04109586225809
1 11.0669676391625
};
\addplot [black]
table {%
0.875 0.0142897494772524
1.125 0.0142897494772524
};
\addplot [black]
table {%
0.875 11.0669676391625
1.125 11.0669676391625
};
\addplot [black]
table {%
1.75 3.18549100061414
2.25 3.18549100061414
2.25 7.11692261739052
1.75 7.11692261739052
1.75 3.18549100061414
};
\addplot [black]
table {%
2 3.18549100061414
2 0.0583230003451795
};
\addplot [black]
table {%
2 7.11692261739052
2 13.0096688959272
};
\addplot [black]
table {%
1.875 0.0583230003451795
2.125 0.0583230003451795
};
\addplot [black]
table {%
1.875 13.0096688959272
2.125 13.0096688959272
};
\addplot [black]
table {%
2.75 2.94197357701769
3.25 2.94197357701769
3.25 6.59471675844934
2.75 6.59471675844934
2.75 2.94197357701769
};
\addplot [black]
table {%
3 2.94197357701769
3 0.0448534489953326
};
\addplot [black]
table {%
3 6.59471675844934
3 12.0679038885493
};
\addplot [black]
table {%
2.875 0.0448534489953326
3.125 0.0448534489953326
};
\addplot [black]
table {%
2.875 12.0679038885493
3.125 12.0679038885493
};
\addplot [black]
table {%
3.75 2.99890013482185
4.25 2.99890013482185
4.25 6.63801822475047
3.75 6.63801822475047
3.75 2.99890013482185
};
\addplot [black]
table {%
4 2.99890013482185
4 0.0565246487173184
};
\addplot [black]
table {%
4 6.63801822475047
4 12.0831665326471
};
\addplot [black]
table {%
3.875 0.0565246487173184
4.125 0.0565246487173184
};
\addplot [black]
table {%
3.875 12.0831665326471
4.125 12.0831665326471
};
\addplot [black]
table {%
4.75 2.95608780445122
5.25 2.95608780445122
5.25 6.48038784857198
4.75 6.48038784857198
4.75 2.95608780445122
};
\addplot [black]
table {%
5 2.95608780445122
5 0.076179914010877
};
\addplot [black]
table {%
5 6.48038784857198
5 11.7556572862352
};
\addplot [black]
table {%
4.875 0.076179914010877
5.125 0.076179914010877
};
\addplot [black]
table {%
4.875 11.7556572862352
5.125 11.7556572862352
};
\addplot [black]
table {%
0.75 4.19487696018194
1.25 4.19487696018194
};
\addplot [black]
table {%
1.75 4.94740247992075
2.25 4.94740247992075
};
\addplot [black]
table {%
2.75 4.51436176056279
3.25 4.51436176056279
};
\addplot [black]
table {%
3.75 4.5975643823576
4.25 4.5975643823576
};
\addplot [black]
table {%
4.75 4.50732190139063
5.25 4.50732190139063
};
\end{axis}
\end{tikzpicture}
        \end{small}
    \end{center}
    \end{subfigure}
    \caption{Box plot of angular distances ($^\circ$) for the five different networks using simulated (a) and real (b) SRIRs. The boxes are drawn from the first to the third quartile. The horizontal line shows the median. The whiskers go from the lowest data still within 1.5 interquartile range (IQR) of the lower quartile to the highest data within 1.5 IQR of the upper quartile.}    \label{fig:box}
\end{figure*}
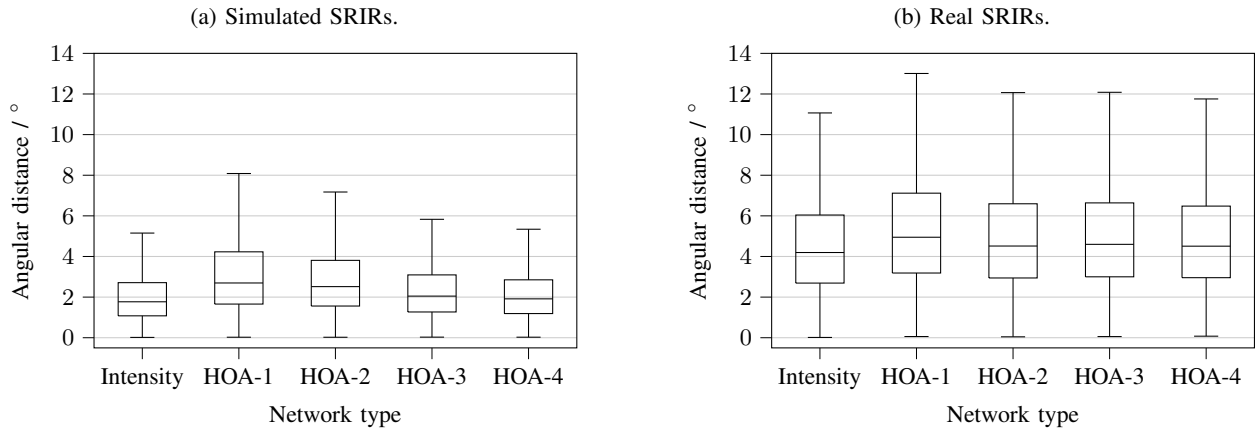
\begin{figure*}[htb]
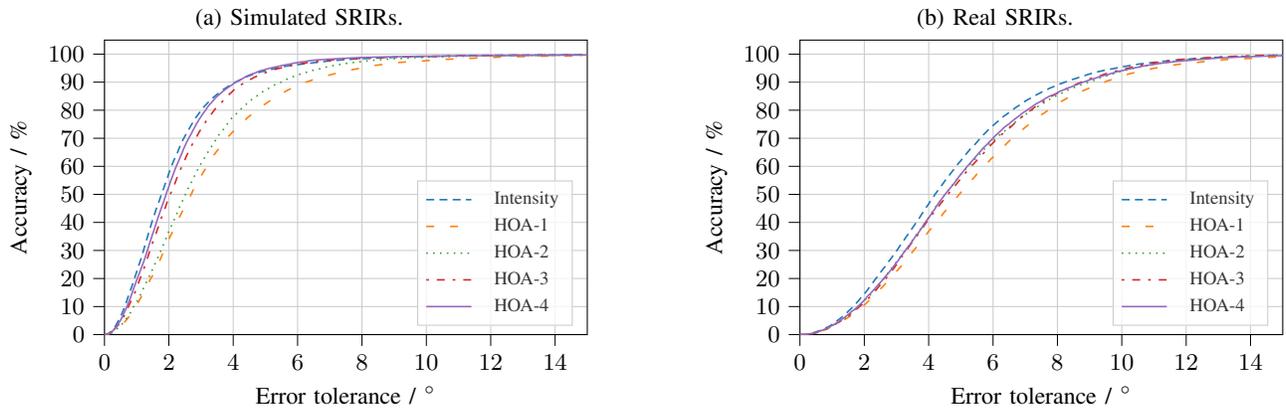

    \begin{subfigure}[htb]{0.49\linewidth}
        \caption{Simulated SRIRs.}\label{fig:dist_sim}
        \vspace*{-10pt}
        \begin{center}
            \begin{small}
                \input{figures/dist_sim}
            \end{small}
        \end{center}
    \end{subfigure}
\hfill
    \begin{subfigure}[htb]{0.49\linewidth}
    \caption{Real SRIRs.}\label{fig:dist_real}
    \vspace*{-10pt}
    \begin{center}
        \begin{small}
            \input{figures/dist_real}
        \end{small}
    \end{center}
\end{subfigure}
\caption{Accuracies of the different networks as a function of the error tolerance for the simulated (a) and real (b) SRIRs.}\label{fig:dist}
\end{figure*}
\setlength\figureheight{6.1cm} 
\begin{figure*}[htb]
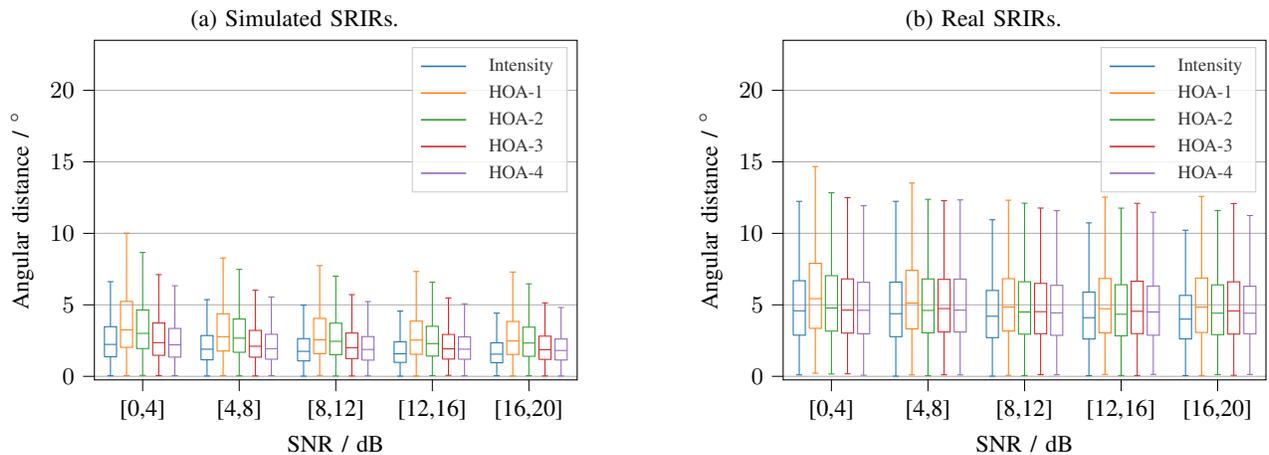

    \begin{subfigure}[htb]{0.495\linewidth}
        \caption{Simulated SRIRs.}\label{fig:sir_box_sim_1_source}
        \vspace*{-10pt}
        \begin{center}
            \begin{small}
                \input{figures/box_sir_average-5_False_1}
            \end{small}
        \end{center}
    \end{subfigure}
    \hfill
    \begin{subfigure}[htb]{0.495\linewidth}
        \caption{Real SRIRs.}\label{fig:sir_box_real_1_source}
        \vspace*{-10pt}
        \begin{center}
            \begin{small}
                \input{figures/box_sir_average-5_True_1}
            \end{small}
        \end{center}
    \end{subfigure}
    \caption{Box plots with angular distances of the different networks for different SNR regions and simulated (a) and real (b) SRIRs.}\label{fig:sir_box}
\end{figure*}
As expected, the results belonging to the simulated SRIRs are overall slightly better than those belonging to the real SRIRs. Nevertheless, all models show a good and reliable generalization ability. 
Alltogether, the results presented in Fig.~\ref{fig:box} and \ref{fig:dist} show that the Intensity-CRNN provides the best localization accuracy on both simulated and real SRIRs. This underlines the statement of Perotin et al. \cite{Perotin.2019} that their intensity features are very well suited for deep learning based DOA estimation

Nevertheless, it can be seen in 
Fig.~\ref{fig:box_sim} and \ref{fig:dist_sim}, that the HOA-$n$-CRNNs perform better with increasing order $n$ on the simulated data. Both the median and the IQR of the angular distance become smaller with each additional order.
In particular, the additional orders of the SH seem to allow a better fine localization. Thus, only about \SI{70}{\%} of the predictions of the HOA-1-CRNN lie within the error tolerance of $4^\circ$, whereas this is the case for about \SI{90}{\%} of the predictions of the HOA-4-CRNN. The rough direction, however, seems already to be well predictable with the HOA-1-CRNN. All considered networks have an accuracy of about \SI{99}{\%} with an error tolerance of $15^\circ$.

However, the results belonging to the real SRIRs in Fig.~\ref{fig:box_real} and \ref{fig:dist_real} show that an improvement of the DOA estimates is only obtained when the order is increased from 1 to 2. The HOA-CRNNs of orders 2 to 4 achieve almost identical results.

In Fig.~\ref{fig:sir_box} the localization accuracy is evaluated as a function of the SNR of the respective speech signal. As expected, the localization becomes more accurate for each model with increasing SNR. For both simulated and real SRIRs, a slight trend can be seen that the advantage of the Intensity-CRNN over the HOA-$n$-CRNN of orders 3 and especially 4 mainly exists at relatively high SNR. In the case of poor SNR between 0 and \SI{4}{dB}, the HOA-4-CRNN performs even sligthly better than the Intensity-CRNN.
Otherwise, the respective order of localization accuracy among the models remains the same.

\section{Conclusion and Outlook}
\label{sec:conclusion}
In this paper we investigated the influence of the order of HOA signals on the accuracy of single-speaker DOA estimation of noisy speech with CRNNs.
We have shown that there is potential in using the additional spatial information of HOA signals for a CRNN-based DOA estimation. However, when evaluated on real data, it has been shown that the advantage of this additional information may possibly be reduced in practice due to effects such as a non-perfect simulation, a limited generalization capability of the models, or additional measurement noise. Rather, it became very clear that it is highly useful and advisable to extract the information present in the signals in a preprocessing step to make it more accessible for the network.
Only in low-SNR conditions a slight improvement of the DOA estimation could be achieved by using fourth order Ambisonics signals comparing to the Intensity-CRNN.

Since the HOA models seem to perform comparatively well in acoustically challenging scenarios, we will also investigate the effect of the Ambisonics order on localization accuracy in multi-speaker DOA estimation scenarios in the future. Also based on the physical motivation and interpretation of the sound Intenisty features, it can be suspected that the higher-order models are superior to the Intensity-CRNN there.


Furthermore, we want to strengthen our results by additional evaluations of our models on more data generated from real SRIRs and also on real recordings. 
In addition, we want to use our presented dataset to estimate additional parameters such as room volume, reverberation time and frequency-dependent absorption and scattering coefficients using HOA signals.



\bibliographystyle{IEEEbib}
\bibliography{refs}

\end{document}